\documentclass[a4paper,11pt]{article}
 
\usepackage{placeins}
\usepackage{jinstpub}
\usepackage{lineno}
\usepackage{upgreek}
\usepackage{isotope}
\usepackage{siunitx}
\usepackage{subcaption}
\usepackage{graphicx} 
\usepackage{isotope}

\title{\boldmath Investigation of Electron Backscattering on Silicon Drift Detectors for the Sterile Neutrino Search with TRISTAN}

\author[a, b, 1]{D.~Spreng \note{Corresponding author.},}
\author[a, b]{K.~Urban,}
\author[c, d]{M.~Carminati,}
\author[a, b]{F.~Edzards,}
\author[c, d]{C.~Fiorini,}
\author[e]{P.~Lechner,}
\author[c, f]{A.~Nava,}
\author[a, b]{D.~Siegmann,}
\author[a, b]{C.~Wiesinger,}
\author[a, b]{and S.~Mertens}

\affiliation[a]{Technical University of Munich, TUM School of Natural Sciences, Physics Department, James-Franck-Str.~1, 85748 Garching, Germany}
\affiliation[b]{Max Planck Institute for Physics, Boltzmannstr. 8, 85748 Garching, Germany}
\affiliation[c]{INFN-Sezione di Milano, Via Celoria 16, 20133 Milano, Italy}
\affiliation[d]{DEIB, Politecnico di Milano, Via Golgi 40, 20133 Milano, Italy}
\affiliation[e]{Halbleiterlabor der Max-Planck-Gesellschaft, Isarauenweg 1, 85748 Garching, Germany}
\affiliation[f]{University of Milano-Bicocca, Piazza della Scienza 3, 20126 Milano, Italy}

\emailAdd{daniela.spreng@tum.de}

\abstract{Sterile neutrinos are hypothetical particles in the minimal extension of the Standard Model of Particle Physics. They could be viable dark matter candidates if they have a mass in the keV range. The Karlsruhe tritium neutrino~(KATRIN) experiment, extended with a silicon drift detector focal plane array~(TRISTAN), has the potential to search for keV-scale sterile neutrinos by measuring the kinematics of the tritium $\upbeta$-decay. The collaboration targets a sensitivity of $10^{-6}$ on the mixing amplitude $\sin^2{\Theta}$. For this challenging target, a precise understanding of the detector response is necessary. In this work, we report on the characterization of electron backscattering from the detector surface, which is one of the main effects that influence the shape of the observed energy spectrum. Measurements were performed with a tandem silicon drift detector system and a custom-designed electron source. The measured detector response and backscattering probability are in good agreement with dedicated backscattering simulations using the \textsc{Geant4} simulation toolkit.}


\begin{document}

\maketitle
\flushbottom

\section{Introduction}
\label{sec:intro}

Sterile neutrinos have gained significant attention in the field of particle physics and astrophysics. In the scope of a minimal extension to the Standard Model of Particle Physics~(SM), they are postulated as the right-handed counterparts to the well-known left-handed neutrinos~\cite{Abazajian_2012}. As a consequence, an additional new neutrino mass eigenstate is introduced. While left-handed neutrinos actively participate in weak interactions, right-handed neutrinos do not interact via any of the fundamental forces described by the SM. Right-handed neutrinos only interact via the gravitational force of the new mass eigenstate and their mixing with the active neutrino flavors. In the following, the new mass eigenstate is referred to as `sterile neutrino'.\par 

The introduction of sterile neutrinos addresses various unresolved questions in cosmology and neutrino physics, contingent upon factors such as their mass $m_{\mathrm{s}}$, the mixing with the active flavor~(described by the mixing amplitude $\sin^2{\Theta}$), and the production mechanism~\cite{Abazajian_2012, Adhikari_2017, Domcke_2021}. By incorporating sterile neutrinos into the framework of the SM, they offer a natural mechanism for mass generation for active neutrinos~\cite{Mohapatra_2007}. Sterile neutrinos are not constrained to a specific mass range, and those in the keV-mass scale would serve as a viable dark matter candidate~\cite{Adhikari_2017}. As such, the mixing amplitude of sterile neutrinos has been subject to stringent constraints from indirect searches and cosmological observations of $10^{-10} < \sin^2{\Theta} < 10^{-6}$ in a mass range of \SI{1}{\kilo\electronvolt} to \SI{50}{\kilo\electronvolt}~\cite{Boyarsky_2009, Narayanan_2000, Seljak_2006, Lee_1977, Boyarsky_2008, Watson_2012}. However, these limits are model-dependent and can be significantly relaxed by several orders of magnitude through modifications to the models of dark matter decay~\cite{Pal_1982}. \par 

By analyzing the $\upbeta$-decay spectrum, it is possible to search for sterile neutrinos independently of cosmological and astrophysical models~\cite{Shrock_1980}. In a $\upbeta^-$-decay, an electron and an electron anti-neutrino $\overline{\nu}_{\mathrm{e}}$ are emitted. The measured energy spectrum of the electrons is a superposition of spectra corresponding to the different neutrino mass eigenstates which compose the neutrino flavor eigenstate $\overline{\nu}_{\mathrm{e}}$. Therefore, a keV-scale neutrino mass eigenstate would display a significantly reduced maximal electron energy compared to the other decay branches. This results in a kink-like signature at $m_\mathrm{s}$ below the endpoint energy $E_0$ of the energy spectrum, accompanied by a global distortion of the spectrum~\cite{Mertens_2019}.\par 

The Karlsruhe Tritium Neutrino (KATRIN) experiment was designed to perform high-precision integral measurements of the tritium $\upbeta^-$-decay spectrum at its endpoint energy of $E_\mathrm{0} =$ \SI{18.6}{\kilo\electronvolt} to test the effective electron antineutrino mass in the sub-eV range~\cite{KATRIN_2021}. The experiment primarily consists of a highly stable gaseous tritium source with an activity of up to \SI{e11}{\becquerel}, a high-resolution spectrometer utilizing the magnetic adiabatic collimation with electrostatic (MAC-E) filter principle, and a detector section. The unique source properties of the KATRIN experiment allow to use it to investigate spectral contributions from sterile neutrinos in the keV range by significantly lowering the MAC-E filter's retarding potential to a constant value and measuring the entire differential $\upbeta$-decay spectrum. As a result, the energy resolution will directly depend on the intrinsic performance of the detector. Furthermore, the detector system needs to manage extremely high count rates. Therefore, after the neutrino mass measurement campaigns are finished at the end of 2025, an upgrade of the detector section is planned, called the TRISTAN detector system. The TRISTAN detector will consist of nine modular 166-pixel silicon drift detector (SDD) arrays and will measure the differential electron energy spectrum with an energy resolution of less than \SI{300}{\electronvolt} FWHM\footnote{FWHM stands for the peak's full width at half maximum.} at \SI{20}{\kilo\electronvolt}~\cite{Mertens_2021}. The spectrometer will continue to be an important tool to ensure adiabatic electron transport, reject low-energy parts of the spectrum dominated by systematic errors, and for calibration. \par 

\begin{figure}[btp]
    \centering
    \includegraphics[width=.7\textwidth]{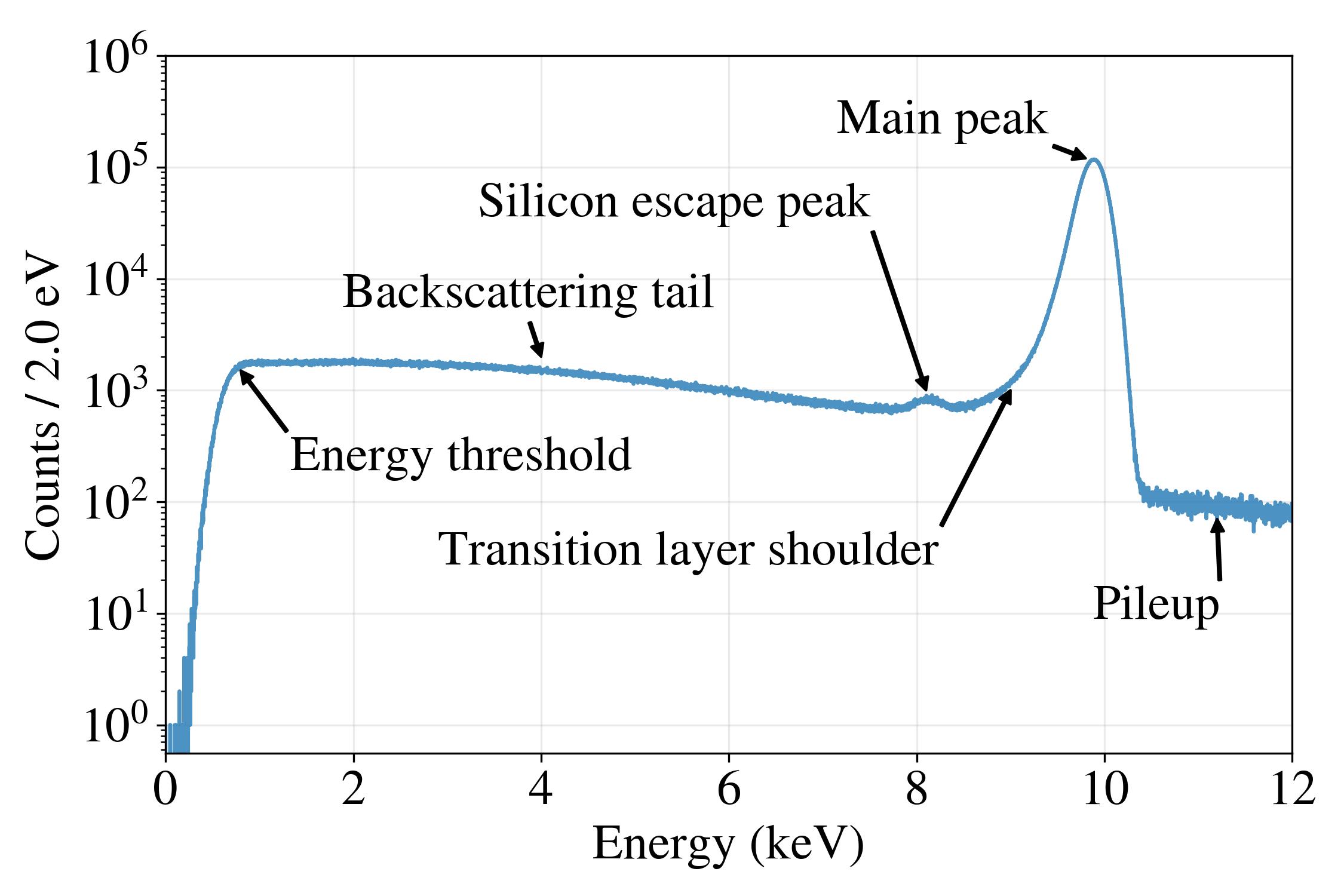}
    \caption{Measured response of the TRISTAN detector to \SI{10}{\kilo\electronvolt} electrons. The detector response differs from purely Gaussian resolution effects. It is modified by various factors, including signal read-out, detector design, and physical effects. The effect of electron backscattering leads to a broad tail from the main peak down to the minimal measurable deposit energy.}
    \label{fig:Figure1}
\end{figure}

With this detector system upgrade, a search for sterile neutrinos with masses up to \SI{18.6}{\kilo\electronvolt} and admixtures down to $10^{-6}$ will be possible, improving current laboratory limits by several orders of magnitude~\cite{Mertens_2019, Abdurashitov_2017, Holzschuh_1999, Holzschuh_2000}. However, for a sterile neutrino search with a mixing-angle sensitivity at the parts-per-million~(ppm) level, a thorough understanding of the detector response is crucial. This includes all experimental influences that alter the measured spectral shape of the tritium spectrum. The detector response for SDDs is primarily dominated by electronic noise, partial charge collection at the entrance window\footnote{In the scope of this work, the entrance window denotes the side through which the electron enters the detector volume.} of the detector~\cite{Gugiatti_2020, Mertens_2021}, and the inevitable effect of electron backscattering, as illustrated in figure~\ref{fig:Figure1}.\par

When an electron undergoes scattering processes inside the detector, its direction of movement randomly changes, potentially resulting in its backscattering towards the entrance window and subsequent exit from the detector. Also, secondary electrons produced in the detector by the incoming electron can escape from the detector volume. Consequently, backscattering is a major source of incomplete charge collection, affecting the measured electron spectrum over the entire energy range. The extent of backscattering depends on the initial energy~$E_{\mathrm{I}}$ and incident angle~$\theta_{\mathrm{I}}$ of the incoming electron. Lower electron energies and higher incident angles lead to a reduced penetration depth relative to the entrance window, increasing the probability of electron escape from the detector. In addition, the presence of electric and magnetic fields in the KATRIN beamline introduces the possibility of backscattered electrons being reflected back towards the detector. Whether the backscattered electron returns to the detector, as well as the temporal and spatial difference between the first and possible second interaction with the detector, depends on the energy and the angle of the backscattered electron.\par 

This work investigates the backscattering properties of the TRISTAN detector system at various initial electron energies and incident angles. To this end, measurements (see section~\ref{sec:Experiment}) and \textsc{Geant4} backscattering simulations (see section~\ref{sec:Simulation}) are performed at different electron energies and angles. By comparing the results (see section~\ref{sec:Comparison}), we assess the ability of \textsc{Geant4} to accurately model backscattering in silicon detectors.

\FloatBarrier
\section{Experimental investigations}
\label{sec:Experiment}
\FloatBarrier

\subsection{Measurement setup}
\begin{figure}[tbp]
    \centering
    \begin{subfigure}[b]{0.517\textwidth}
        \centering\includegraphics[width=0.99\linewidth]{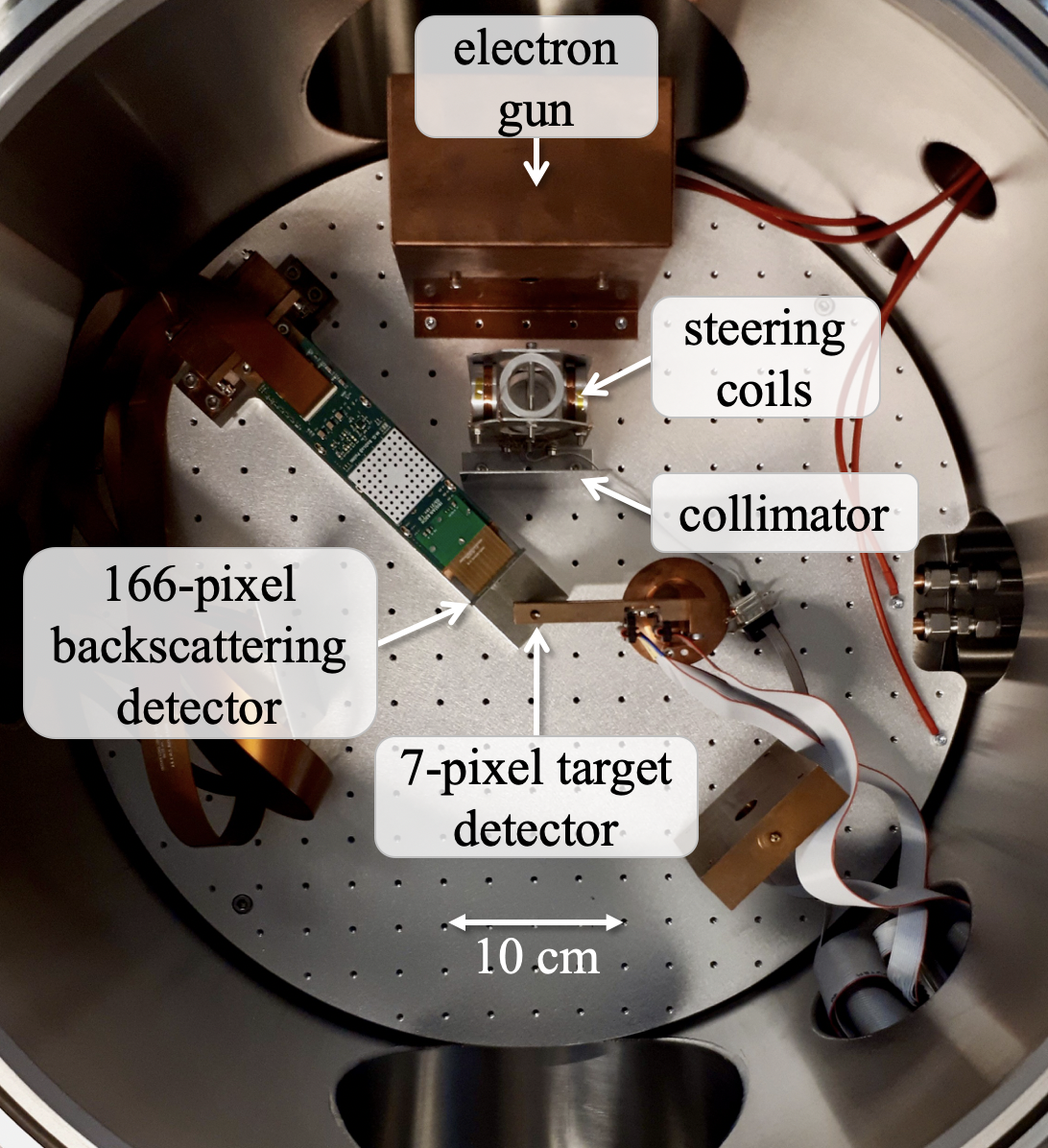}
        \subcaption{Top view}
        \label{fig:Figure2a}
    \end{subfigure}
    \begin{subfigure}[b]{0.476\textwidth}
        \centering\includegraphics[width=0.99\linewidth]{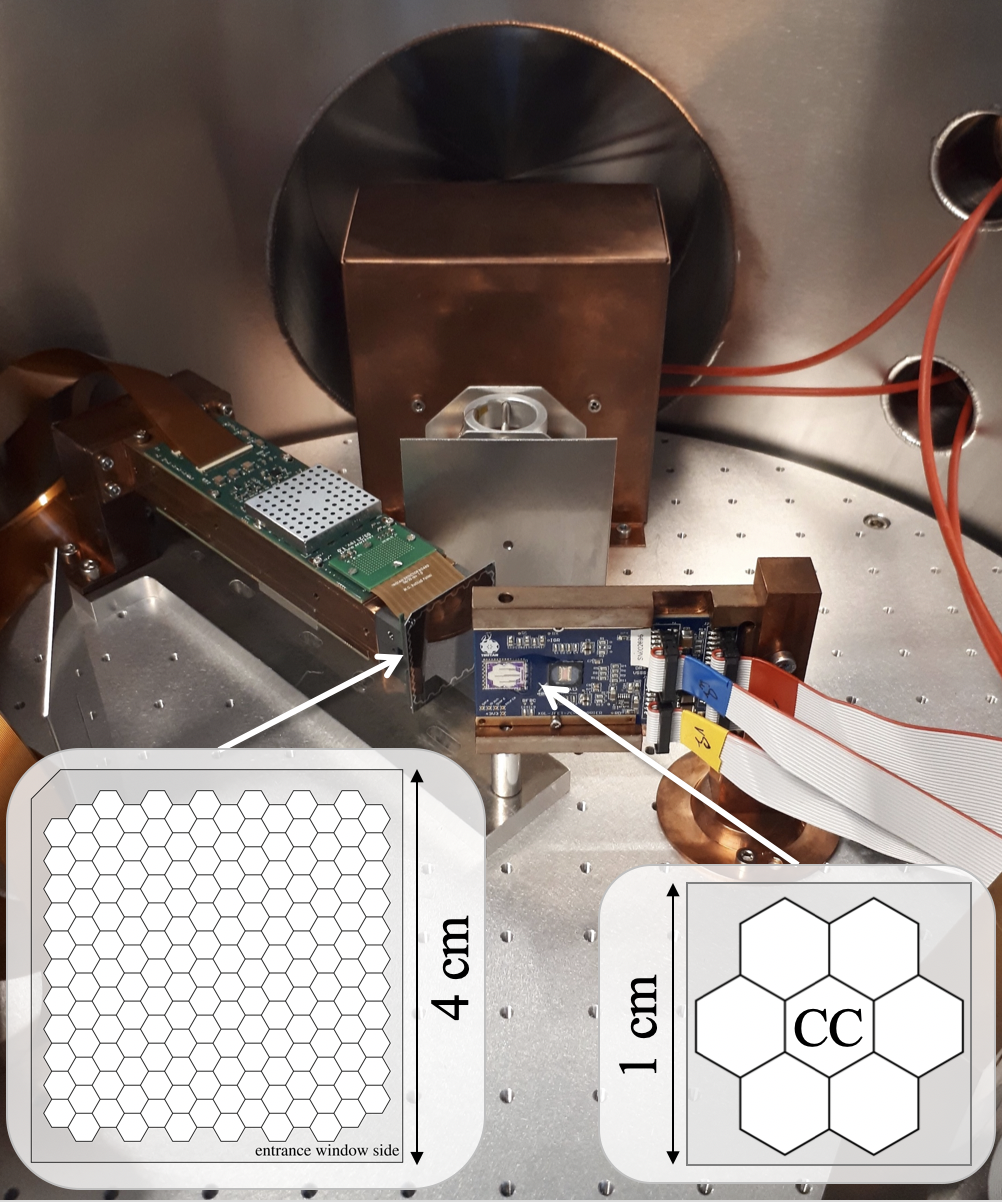}
        \subcaption{Oblique view}
        \label{fig:Figure2b}
    \end{subfigure}
	\caption{In-vacuum setup. The experimental setup is depicted from a top view~(figure~\ref{fig:Figure2a}) and an oblique view~(figure~\ref{fig:Figure2b}). The electron gun is positioned inside a copper shielding. It emits a mono-energetic electron beam directed towards the 7-pixel target detector. The electrons hit the target detector through a rectangular hole in the PCB. The 166-pixel backscattering detector faces the target detector. Following the electron gun, steering coils and a collimator are placed to deflect and shield misaligned electrons. The entire setup is mounted onto a cooling plate installed inside the vacuum chamber.}
	\label{fig:Figure2}
\end{figure}

Two TRISTAN detectors are utilized to measure electron backscattering:~a 7-pixel prototype detector as an active target, in the following called target detector, and a 166-pixel detector module~\cite{Siegmann_2024} to detect the backscattered electrons, in this work called backscattering detector. The pixels are hexagons with a circumscribed diameter of approximately \SI{3.3}{\milli\meter} and a thickness of \SI{450}{\micro\meter} arranged in a honeycomb structure. Both detectors are mounted on copper holding structures and positioned on the cooling plate of a cylindrical vacuum chamber. Electrons from a custom-designed electron gun~\cite{Urban_2024} are directed towards the target detector. The geometry of the experimental setup is depicted in figure~\ref{fig:Figure2}.\par

The electron gun comprises an electrically heated tantalum wire which emits thermal electrons. A negative high voltage up to \SI{10}{\kilo\volt} is applied between the wire and a grounded copper electrode to accelerate these electrons. The uncertainty of the electron's kinetic energy was not measured yet, but it is expected to be in the order of \SIrange{1}{10}{\electronvolt} and is, therefore, negligible considering the detector energy resolution of about \SIrange{200}{300}{\electronvolt} in the measured electron energy range. The electrons leave the electron gun through a hole in the copper electrode. The rate of electrons reaching the target detector can be adjusted by the heating current in the wire between about \SIrange[range-phrase=~and~]{1}{10}{\kilo cps}\footnote{Here, cps abbreviates counts per second.}. \par 

\begin{figure}[tbp]
    \centering
    \begin{subfigure}[b]{0.49\textwidth}
        \centering\includegraphics[width=0.6\linewidth]{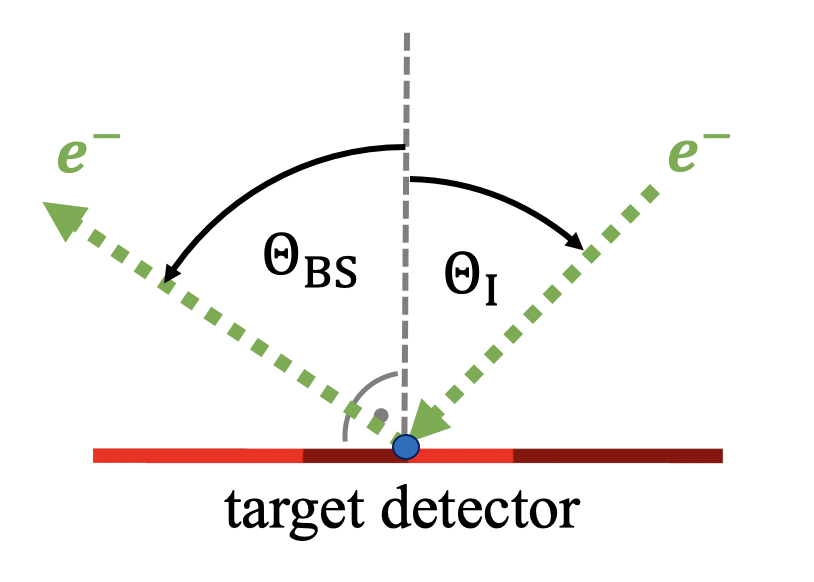}
        \subcaption{Target-detector-rotation and angle illustration}
        \label{fig:Figure3a}
    \end{subfigure}
    \begin{subfigure}[b]{0.49\textwidth}
        \centering\includegraphics[width=0.65\linewidth]{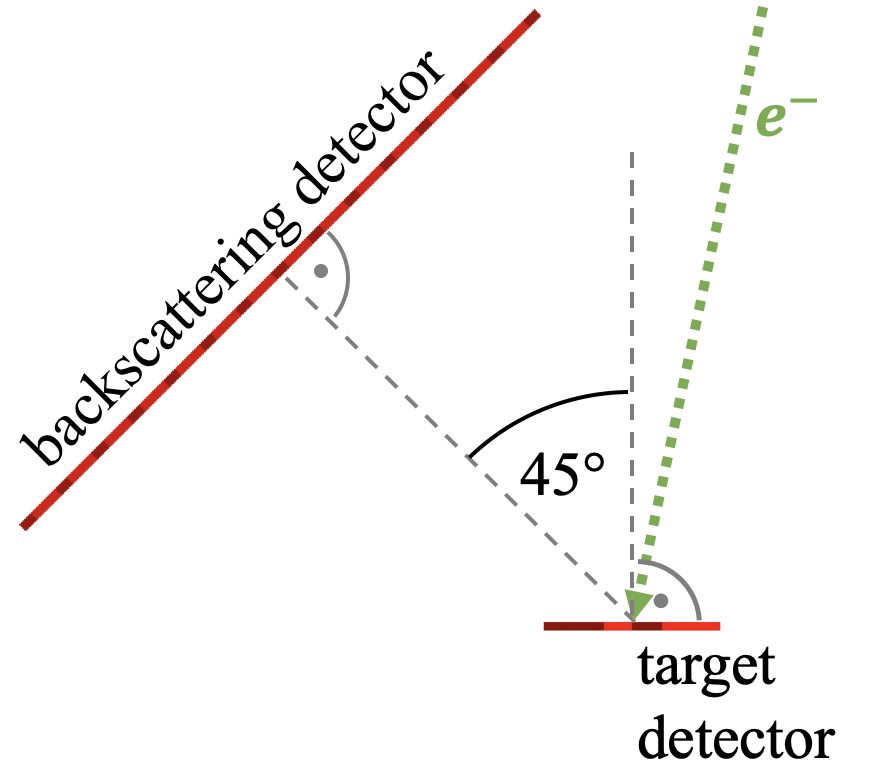}
        \subcaption{Take-off angle illustration}
        \label{fig:Figure3b}
    \end{subfigure}
	\caption{Geometry illustration. Figure~\ref{fig:Figure3a} illustrates the target detector (red) from a top view. The axis of rotation is denoted by a blue dot. Green arrows trace the path of the electron beam, with incident angle~$\Theta_{\mathrm{I}}$ and azimuthal backscattering angle~$\Theta_{\mathrm{BS}}$ indicated. Figure~\ref{fig:Figure3b} shows the top view of both the target and backscattering detector (red). The path of the electron beam is depicted in green, and the take-off angle (fixed at 45°) is marked.}
	\label{fig:Figure3}
\end{figure}

The resulting mono-energetic electron beam is directed at the central pixel (CC) of the 7-pixel target detector. To investigate various incident angles between the electron beam and the target detector, the PCB with the target detector can be rotated around an aluminium post. The axis of rotation as well as the incident angle are illustrated in figure~\ref{fig:Figure3a}. The setup provides the flexibility to choose any incident angle between \SI{0}{\degree} and \SI{60}{\degree}.\par 

The 166-pixel backscattering detector faces the target detector to detect the backscattered electrons. Its positioning was deliberately chosen to ensure maximum angular coverage within a single measurement without a direct line of sight between the electron gun and the backscattering detector. The backscattering detector is fixed on an aluminum mounting structure, which enables a rotation around the same axis as the target detector. It maintains a fixed distance between the detectors but allows adjustments to the take-off angle, which is set to \SI{45}{\degree} for all measurements (see figure~\ref{fig:Figure3b}). As the target detector rotates to change the incident angle of the electron beam, the backscattering detector is also rotated to preserve the same orientation relative to the target detector. Some pixels of the backscattering detector exhibit noisy spectra, encounter connection issues, or are partially shaded by the printed circuit board (PCB) of the target detector in the experiment. Therefore, 43 pixels of the backscattering detector can not be considered for the later comparison of the data with the simulation.\par

To ensure proper alignment of the electron beam with the central pixel of the target detector, steering coils are inserted downstream from the electron gun, enabling the beam to be magnetically deflected in both horizontal and vertical directions. The electron beam diameter is reduced with a collimator inserted after the steering coils to safeguard the backside of the backscattering detector and the PCB of the target detector. The collimator is a thin aluminum plate with a small hole of \SI{3}{\milli\meter}, resulting in a beam diameter of about \SI{9}{\milli\meter} at the target detector plane. All parts were designed and fabricated such that the electron gun hole, the steering coils, the collimator and the central pixel of the target detector are aligned to each other. The alignment was verified by observing the electron rate across all seven pixels of the target detector, confirming that the central pixel receives the highest count rate. For later comparisons with simulated electron energy spectra, only events in the central pixel of the target detector are considered.\par

Both detectors are calibrated using an \isotope[55]{Fe} source. Afterwards, a series of nine measurements is conducted to explore different configurations of the initial energy~$E_{\mathrm{I}}$ and incident angles~$\Theta_{\mathrm{I}}$ of the incoming electrons at the target detector. Each combination for $E_{\mathrm{I}} \in \{ \SI{5}{\kilo\electronvolt}, \SI{7.5}{\kilo\electronvolt}, \SI{10}{\kilo\electronvolt} \}$ and $\Theta_{\mathrm{I}} \in \{ \SI{0}{\degree}, \SI{31}{\degree}, \SI{59}{\degree} \}$ is investigated. Count rates ranging from \SIrange[range-phrase=~to~]{2}{7}{\kilo cps} are observed in the central pixel of the target detector. That variation in rate occurs due to small instabilities of the electron gun and increased irradiation of the surrounding pixels at higher incident angles. For the data acquisition, three synchronized 64-channel CAEN VX2740 digitizer cards are used, which perform a full waveform digitization of each pixel in parallel. An online trapezoidal filter is applied to reconstruct the energy and time information of each event in every pixel, which are used to perform coincidence analysis. The energy, pixel number, and timestamp are recorded for every detected event in either of the detectors.\par 

\subsection{Coincidence analysis}

In the experiment, all pixels of the target detector are exposed to the electron beam due to its angular spread. Since the measured backscattering properties are affected by variations in the incident angle at the target detector, it is important to extract the backscattered electrons originating only from the central pixel of the target detector. As both detectors are active, events in the backscattering detector can be correlated to events in the target detector through their timestamp. This event correlation can be accomplished with a coincidence analysis. This means only events where the time difference between the event in the central pixel of the target detector and the event in the backscattering detector is shorter than a given time window are selected. In the following, a time window of \SI{500}{\nano\second} is chosen, as it is approximately the maximal electron drift time in a pixel~\cite{Forstner_2023}. The time the electron needs to propagate from one detector to the other is in the order of \si{\nano\second} or below for the observed energies and the detector distances and is therefore negligible.\par 

\begin{figure}[btp]
    \centering\includegraphics[width=0.7\linewidth]{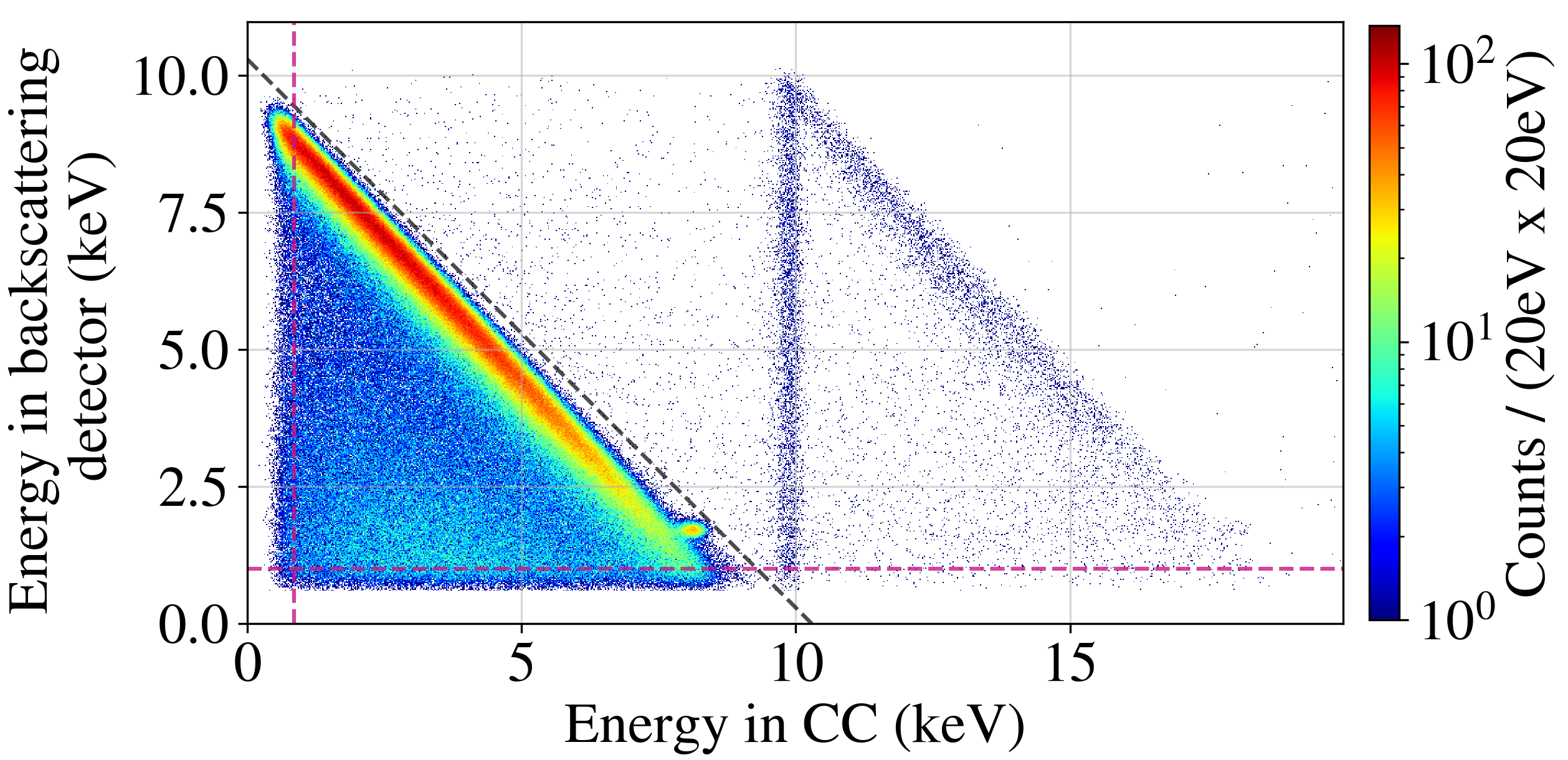}

	\caption{2D-Histogram of identified coincidence events for initial electron energies of \SI{10}{\kilo\electronvolt} and an incident angle of \SI{0}{\degree}. On the x-axis, the histogram displays the energy detected in the central pixel~(CC) of the target detector, while the y-axis represents the energy of the corresponding coincidence event detected in any pixel of the backscattering detector. The black dashed line marks the upper energy limit on the sum of the energies of both events. The purple dashed vertical and horizontal lines indicate the chosen energy thresholds for coincidence events. Both conditions on the coincidence event energy are applied in the subsequent analysis of energy spectra.}
	\label{fig:Figure4}
\end{figure}

In figure~\ref{fig:Figure4}, the energies of all coincidence events detected are depicted in a 2D-histogram exemplary for initial electron energies of \SI{10}{\kilo\electronvolt} and an incident angle of \SI{0}{\degree}. As anticipated, the sum of the energies of both events is nearly equal to the initial electron energy, resulting in a diagonal line in the plot. Due to energy losses in the transition layers of the detectors, the full initial energy cannot be reconstructed. Events falling below this diagonal are either random coincidences or cases where not all backscattered electrons were detected. The second diagonal, which is parallel to the first but shifted by approximately \SI{10}{\kilo\electronvolt} towards higher energies in the target detector, arises from coincidence between a backscattered electron in the backscattering detector and a pileup event in the target detector. The energy thresholds of both detectors are visible as almost blank stripes on the left and bottom side of the histogram. Events besides the explained structures, including the vertical line at \SI{10}{\kilo\electronvolt}, are from random coincidence.

\begin{figure}[tbp]
    \centering
    \begin{subfigure}[b]{0.49\textwidth}
        \centering\includegraphics[width=0.99\linewidth]{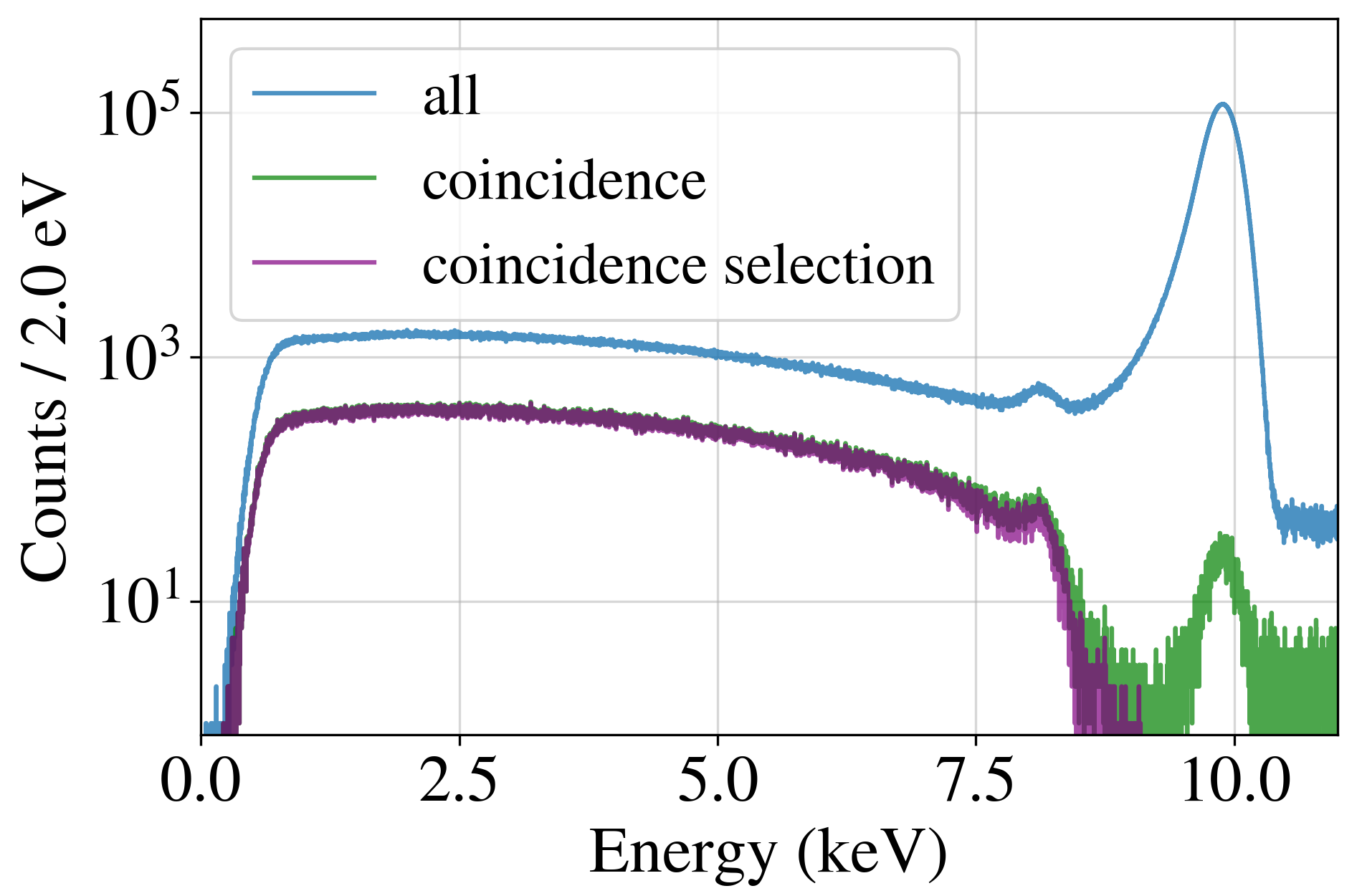}
        \subcaption{Target detector}
        \label{fig:Figure5a}
    \end{subfigure}\hfill
    \begin{subfigure}[b]{0.49\textwidth}
        \centering\includegraphics[width=0.99\linewidth]{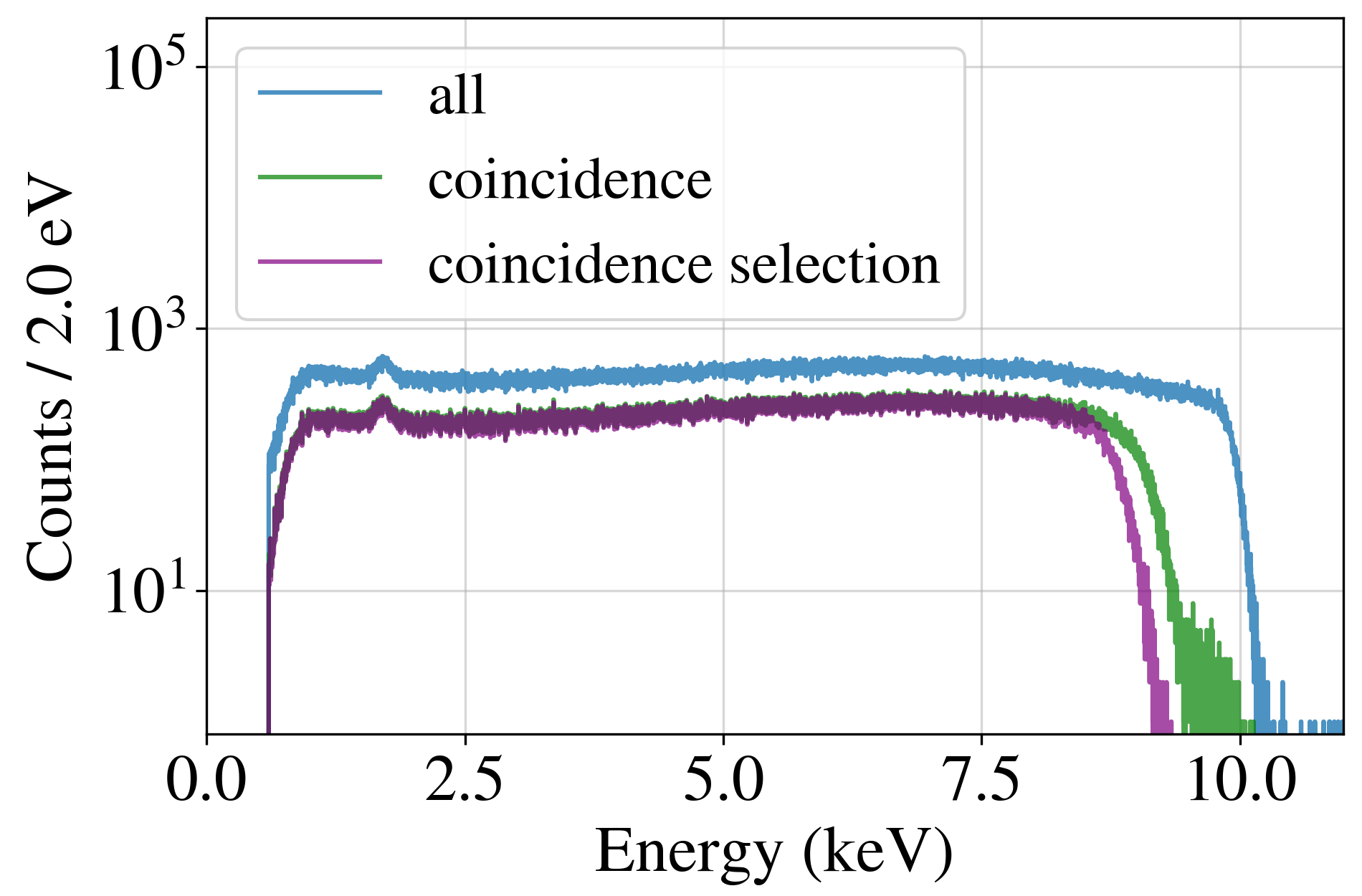}
        \subcaption{Backscattering detector}
        \label{fig:Figure5b}
    \end{subfigure}
	\caption{Typical electron energy spectra with and without coincidence analysis. The energy spectrum of the incoming electron measured in the central pixel of the target detector~(figure~\ref{fig:Figure5a}) and the backscattered electrons measured in any pixel of the backscattering detector~(figure~\ref{fig:Figure5b}) are shown before and after applying the coincidence cut. In green, the spectra are shown for all coincidence events detectable. In purple, the spectra are calculated after additionally applying lower and upper limits on the energies of the coincidence events. The spectra are shown for an initial electron energy of \SI{10}{\kilo\electronvolt} and an incident angle of \SI{0}{\degree}.}
	\label{fig:Figure5}
\end{figure}

Figure~\ref{fig:Figure5} illustrates a typical energy spectrum of the incoming electrons measured by the central pixel of the target detector and the backscattered electrons measured in all pixels of the backscattering detector before and after applying the coincidence analysis. In the target detector, the energy spectrum of the coincidence events mainly resembles the shape of the backscattering tail observed in the overall measured energy spectrum of the target detector. Hence, the coincidence analysis enables the identification and extraction of the signature of backscattered electrons in the target detector. The height of the backscattering tail before and after the coincidence analysis differs due to the backscattering detector's incomplete angular coverage of the space. \par

The shape of the coincidence spectrum measured in the backscattering detector also resembles the shape of the total recorded backscattering spectrum. The fact that not all events in the backscattering detector are also coincidence events can be explained by the relative number of electrons impacting the surrounding pixels or insensitive area rather than the central pixel of the target detector.\par  

For both detectors, the energy spectra of coincidence events exhibit a truncation at higher energies due to the energy threshold of the detectors. If one of the events falls below the energy threshold of either detector, the other event cannot be identified as a coincidence event. This situation arises in cases of elastic backscattering, where the initial electron is reflected off the target detector surface, or inelastic backscattering in the partially insensitive detection area at the entrance window. Additionally, secondary backscattered electrons have predominantly low energies and are likely to fall below the energy threshold of the backscattering detector. In the coincidence spectrum of the target detector, shown in figure~\ref{fig:Figure5a}, the event rate at the full energy (here \SI{10}{\kilo\electronvolt}) is strongly reduced due to the energy threshold of the backscattering detector. Remaining events above $E = E_\mathrm{I} - E_\mathrm{thres}$ are due to random coincidences. Analogously, the energy threshold of the target detector leads to a reduction of the maximally detected coincidence event energy in the backscattering detector (figure~\ref{fig:Figure5b}). Therefore, maintaining a low energy threshold is crucial for investigating backscattering.\par 

To keep simulation and measurements comparable, artificial step-like energy thresholds are applied instead of modeling the detailed energy filtering of the data acquisition system in the experiment. This simplifies the process, as modeling the exact more gradual energy cutoff at low energies (as marked in figure~\ref{fig:Figure1}) would require simulating the entire readout chain and extracting the electron energy from waveforms. Consequently, the energy thresholds for further coincidence selection were fixed at higher energies than the actual energy cutoff in the experiment, as indicated in figure~\ref{fig:Figure4}. For an event in the target detector, the corresponding event in the backscattering detector must have an energy greater than \SI{1}{\kilo\electronvolt}, and for an event in the backscattering detector, the corresponding event in the target detector must not be below \SI{0.85}{\kilo\electronvolt}.\par

As stated above, all electrons detected above the truncation (including the \SI{10}{\kilo\electronvolt} peak in the coincidence spectrum of the target detector depicted in green) are caused by random coincidence; thus, an incoming and a backscattered electron just by chance simultaneously hit the detectors without causal relation. In the experiment, it is further possible for a coincidence to occur between a pileup event in the target detector and an event in the backscattering detector. To reduce both effects, an upper energy limit is imposed on the measured total energy of both coincidence events for the later comparison with the simulation. The total energy of the coincidence event in the target detector, combined with the energy of the coincidence event in the backscattering detector, must not exceed the initial electron energy, accounting for the energy resolutions of both detectors. This upper energy limit is illustrated in figure~\ref{fig:Figure4}.\par

Both those conditions on the minimal and maximal energy of the coincidence events, dealing with the energy threshold and random coincidences, are from now on applied in order to compare the experiment with the more simplistic simulation. The resulting coincidence energy spectra are depicted in purple in figure~\ref{fig:Figure5}.

\FloatBarrier
\section{Geant4 backscattering simulations}
\label{sec:Simulation}
\FloatBarrier

The experimental setup is implemented in a stand-alone application based on the \textsc{Geant4} simulation framework. The \textsc{G4EmStandardPhysicsSS} physics list is utilized for all simulations~\cite{Geant4PhysicsListGuide}\footnote{A production threshold of \SI{100}{\electronvolt} is applied.}. Unlike the default physics list for low-energy interactions, it treats individual scattering events separately rather than combining them into a multiple scattering process. While this approach is computationally more intensive, it is preferred for backscattering studies conducted in this work. This work does not include uncertainties arising from selecting a particular \textsc{Geant4} physics list.\par
In total, three different detector setups were implemented to pursue three different goals: 1) the determination of parameters describing the entrance window of the detector, 2) the simulation of the incident and backscattering energy spectra in a realistic experimental setup, and 3) the estimation of the total efficiency to detect backscattering electrons, given the geometrical configuration. Each setup and its purpose will be described in the following.\par

\begin{figure}[tbp]
    \centering
    \begin{subfigure}[b]{0.45\textwidth}
        \centering\includegraphics[width=0.95\linewidth]{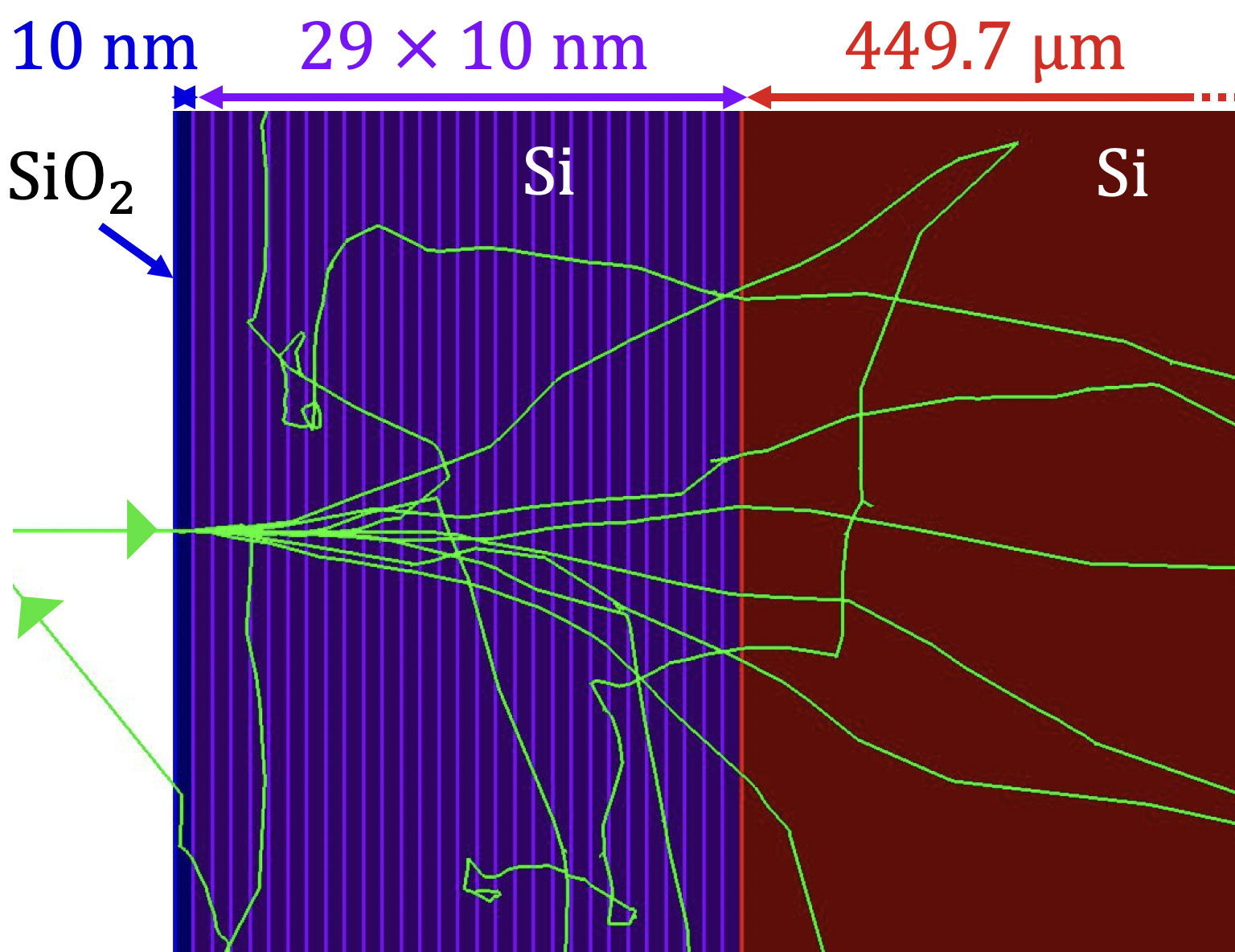}
        \subcaption{Zoom on the entrance window for setup 1}
        \label{fig:Figure6a}
    \end{subfigure}\hfill
    \begin{subfigure}[b]{0.53\textwidth}
        \centering\includegraphics[width=0.85\linewidth]{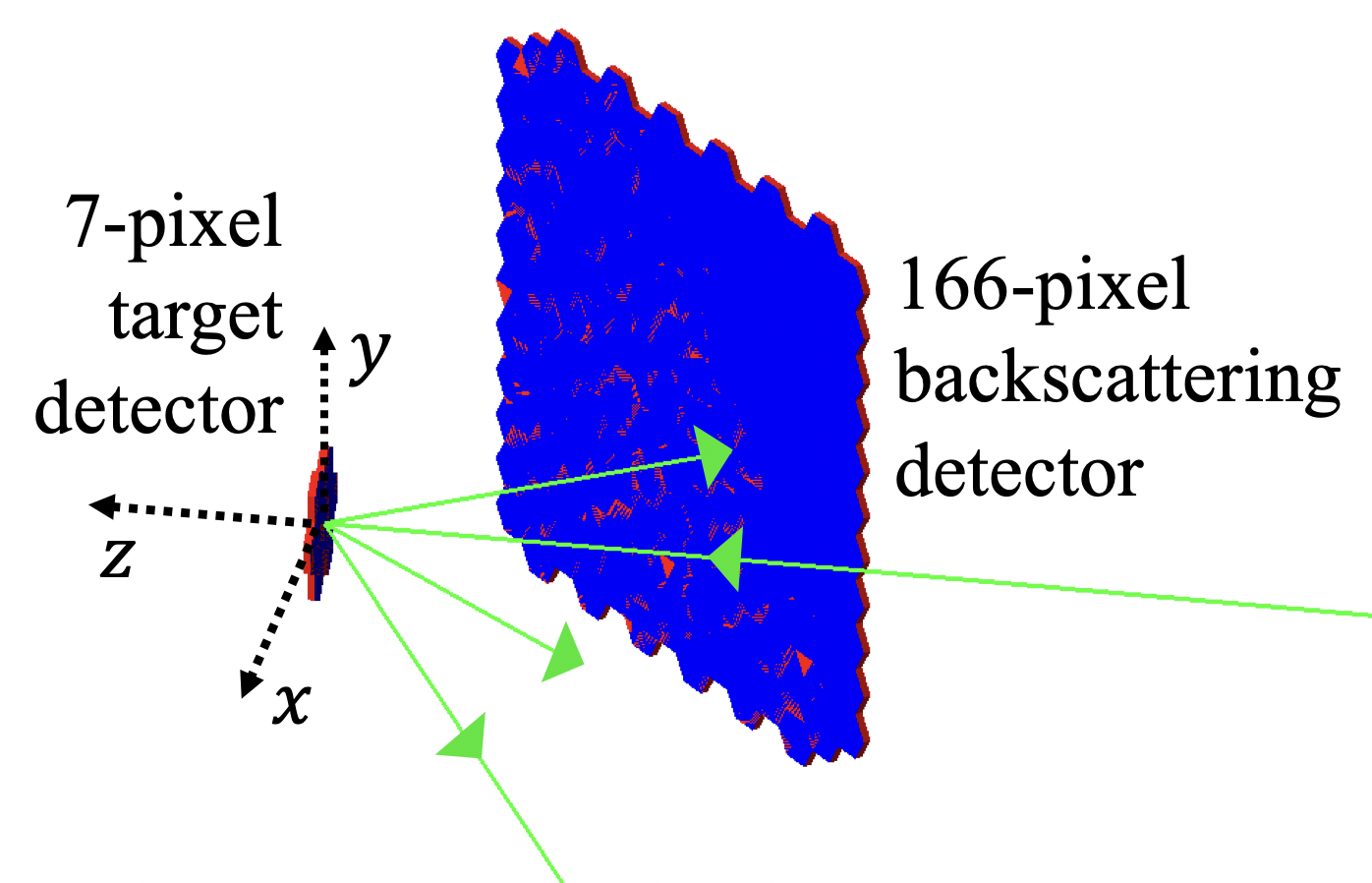}
        \subcaption{Side view of setup 2}
        \label{fig:Figure6b}
    \end{subfigure}\hfill
	\caption{Experimental setup as implemented in the \textsc{Geant4} simulation toolkit. In figure~\ref{fig:Figure6a}, the layered implementation of one pixel to fit the charge collection efficiency at the entrance window is depicted. The pixel consists of a thin silicon-dioxide layer (blue), 29 silicon layers (alternating shades of purple) and a silicon bulk (red). The second implemented setup is visualized from a side view in figure~\ref{fig:Figure6b}. The arrangement of hexagon-shaped pixels, consisting of silicon (red) and a thin silicon-dioxide layer (blue), matches the experimental configuration. Dotted black arrows indicate the coordinate system. In both graphics, exemplary electron paths are displayed using green arrows.}
	\label{fig:Figure6}
\end{figure}

\subsection{Entrance window parameters} 
\label{sec:Sim-EWP}
In the simulation, the silicon detector pixels are generated as hexagons made of silicon with a \SI{10}{\nano\meter} silicon-dioxide layer on the entrance window side\footnote{For the TRISTAN detectors, a silicon-dioxide layer is intentionally added to prevent it from growing naturally. It is manufactured with a controlled homogeneous thickness of \SIrange{8}{10}{\nano\meter}~\cite{Siegmann_2019, Mertens_2021}.}. To account for incomplete charge collection at the entrance window side, as it occurs in the experiment due to the doping profile and the resulting electric field configuration, each individual energy deposition in a pixel has to be weighted by the charge collection efficiency as defined in equation~\ref{eqn:CCE}~\cite{Nava_2021}.

\begin{equation}
    \mathrm{CCE}(z; DL, p_{1}, \lambda) = \left\{
    \begin{array}{ll}
        0, & \hspace{1cm} z<DL \\
        1+(p_{1}-1)\cdot\exp{\left(-\frac{z-DL}{\lambda}\right)}, & \hspace{1cm} z>DL
    \end{array}
    \right.
    \label{eqn:CCE}
\end{equation}

The charge collection efficiency~(CCE) of a detector pixel depends on the position~$z$ of the interaction point relative to the entrance window. The formula indicates that the detector is insensitive within a dead layer of thickness~$DL$. Beyond this layer, the charge collection efficiency follows a step-like increase to a value $p_1$ and then exponentially rises based on the effective transition layer thickness~$\lambda$.\par

To determine the CCE parameters, a setup consisting of a single pixel is implemented in \textsc{Geant4}. At the entrance window side, the pixel consists of 30 layers of \SI{10}{\nano\meter} thickness each, as depicted in figure~\ref{fig:Figure6a}. The first layer is composed of silicon-dioxide while the remaining layers and an adjoining bulk layer of \SI{449.7}{\micro\meter} thickness are made of silicon. The simulation records the accumulated energy depositions for each electron in each layer and the bulk material. Each electron is generated individually and hits exactly the center of the pixel, i.e. charge sharing between pixels or charge losses at the boundaries are neglected in the simulation. A total of 10 million electrons are simulated for each combination of the initial electron energy~$E_\mathrm{I}$ and incident angle~$\Theta_\mathrm{I}$. This is approximately in the order of magnitude of measured counts in the central pixel of the 7-pixel target detector in the experiment.\par

\begin{figure}[btp]
    \centering\includegraphics[width=0.65\linewidth]{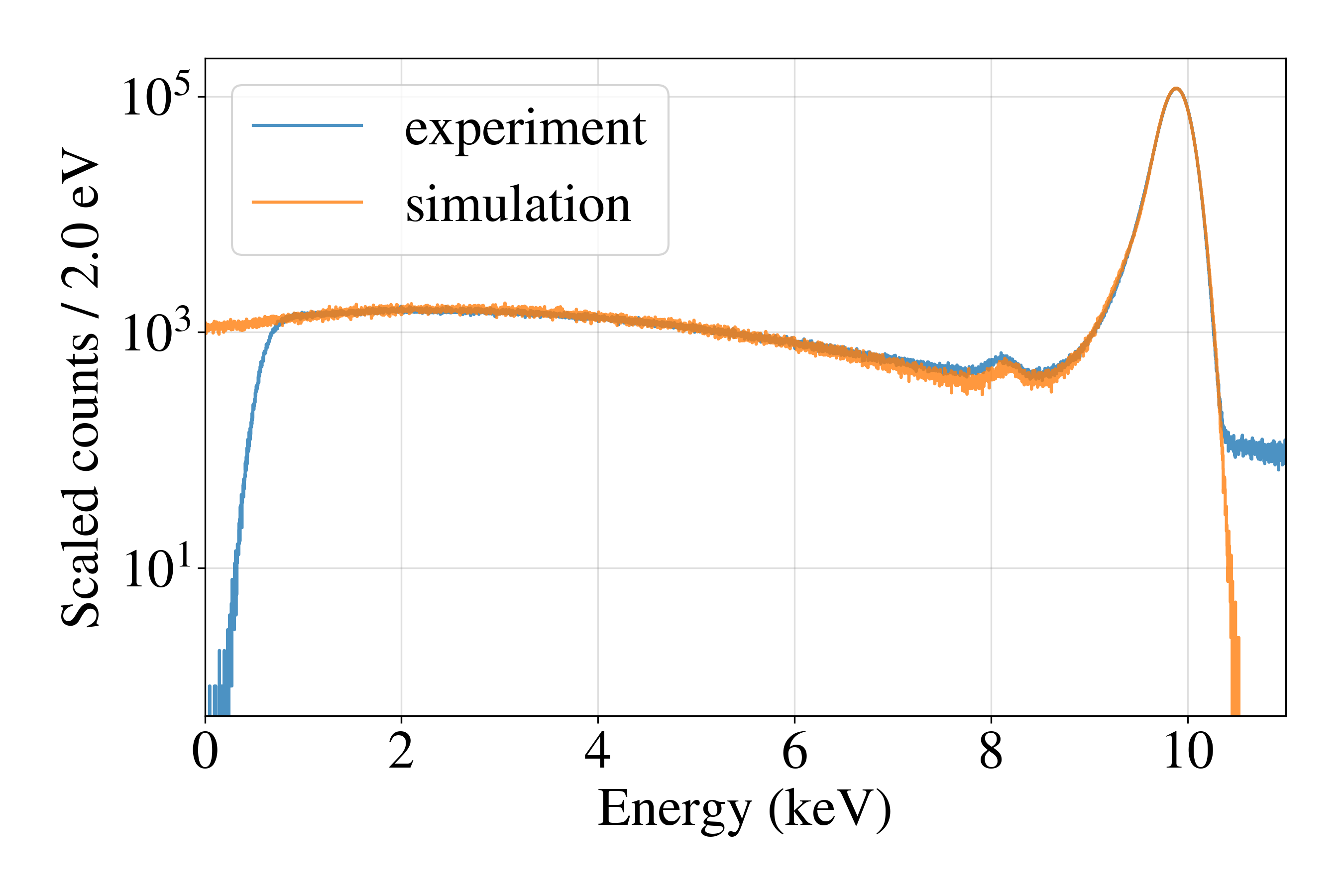}

	\caption{Exemplary fit result ($E_\mathrm{I}=$\SI{10}{\kilo\electronvolt}, $\Theta_\mathrm{I}=$\SI{0}{\degree}). The experimental electron energy spectrum recorded by the CC pixel of the 7-pixel target detector after being corrected for charge sharing between pixels is shown in blue. The simulated data using a layered detector model (setup 1) is fitted to the experimental spectrum to deduce the CCE parameters. The resulting spectrum (shown in orange) is scaled by an amplitude~$A$ to match the total number of measured electrons in the experiment. The amplitude was treated as a fit parameter and amounts here to about $2.6$.}
	\label{fig:Figure7}
\end{figure}

For each setting, the simulation is fitted to the measured data recorded in the central pixel of the 7-pixel target detector with the parameters of the CCE as free parameters for each setting. The fitting process also considers possible energy miscalibrations of the detector system, electronic noise of the read-out chain in the experiment, and the difference in the number of generated incoming electrons between simulation and experiment. An exemplary fit result can be seen in figure~\ref{fig:Figure7}. While the fit does not describe the transition layer shoulder of the main peak and the silicon escape peak highly accurately, the backscattering tail shows a sufficient agreement with the experimental spectrum. As a result, the fit is assumed to perform sufficiently well to deduce approximate values for the CCE parameters for the investigation of the backscattering tail.\par

The average of the CCE parameters obtained for each setting is used in the actual backscattering simulation, which will be described in the following subsection. An average dead layer thickness of $DL = 10.7^{+2.8}_{-6.7}$~\si{\nano\meter} is evaluated, which aligns with the design value of \SIrange{8}{10}{\nano\meter} for the thickness of the insensitive silicon-dioxide layer. The average value for $p_{1}$ is $0.856^{+0.042}_{-0.091}$. Hence, $85.6$~\si{\percent} of the charge deposited right after the dead layer is detected. The mean effective transition layer thickness amounts to $\lambda = 75.7^{+25.1}_{-26.2}$~\si{\nano\meter}. The uncertainties here are deduced as the minimal and maximal values obtained for each parameter comparing the nine fits performed for the nine combinations of $E_\mathrm{I}$ and $\Theta_\mathrm{I}$ settings.

\subsection{Backscattering simulation} 
In a second step, the full experimental setup, including the 7-pixel target and 166-pixel backscattering detector are simulated in \textsc{Geant4}, see figure~\ref{fig:Figure6b}. In this setup, the pixels are not layered anymore. They only consist of a \SI{10}{\nano\meter} silicon-dioxide layer on the entrance window and a silicon bulk. For each event, the energy depositions within the pixel are weighted by the average charge collection efficiency as derived from the simulation described in section~\ref{sec:Sim-EWP}, summed up and recorded alongside the event and pixel number. Additional structures, such as the PCBs and copper holding structures, can be neglected in the simulation since, due to the requirement that both SDDs recorded a signal (coincidence cut), events hitting the insensitive PCB are not recorded. The read-out chain and data acquisition system are not implemented. The simulated energy spectra are convoluted with a Gaussian function to emulate a realistic energy resolution.\par  

In the simulation, each electron is generated individually and precisely hits the central pixel of the 7-pixel target detector at its center. In this work, the actual beam profile is assumed to be negligible. As in the measurement, only electrons coincident between the 7-pixel and 166-pixel detector are considered in the simulation. As for the first set of simulations with the former setup, a total of 10 million electrons are simulated for each combination of the initial electron energy~$E_\mathrm{I}$ and incident angle~$\Theta_\mathrm{I}$. This set of simulations is used to compare the simulated energy spectra of the backscattered electrons with the ones recorded by the backscattering detector in the experiment (see section~\ref{sec:ComparisonSpectra}).\par

\subsection{Detection efficiency}
Due to the energy threshold of both detectors and the geometrical coverage of the 166-pixel backscattering detector, not all incident and backscattered electrons are recorded. To quantify these ineffiencies another simulation is performed. In this simulation, the total number of backscattered electrons is measured and compared to the number that was recorded with the 166-pixel detector of the former simulated setup. With the help of this simulation, the fraction of measured backscattering electrons can be converted into a total backscattering probability. As for the former simulations, simulations are performed with this setup generating 10 million electrons for each combination of the initial electron energy~$E_\mathrm{I}$ and incident angle~$\Theta_\mathrm{I}$. \par 

\FloatBarrier
\section{Results and comparison}
\label{sec:Comparison}

\FloatBarrier
\subsection{Electron energy spectra}
\label{sec:ComparisonSpectra}
\FloatBarrier 

\begin{figure}[btp]
\centering
    \begin{subfigure}[b]{0.475\textwidth}
        \centering\includegraphics[width=\textwidth]{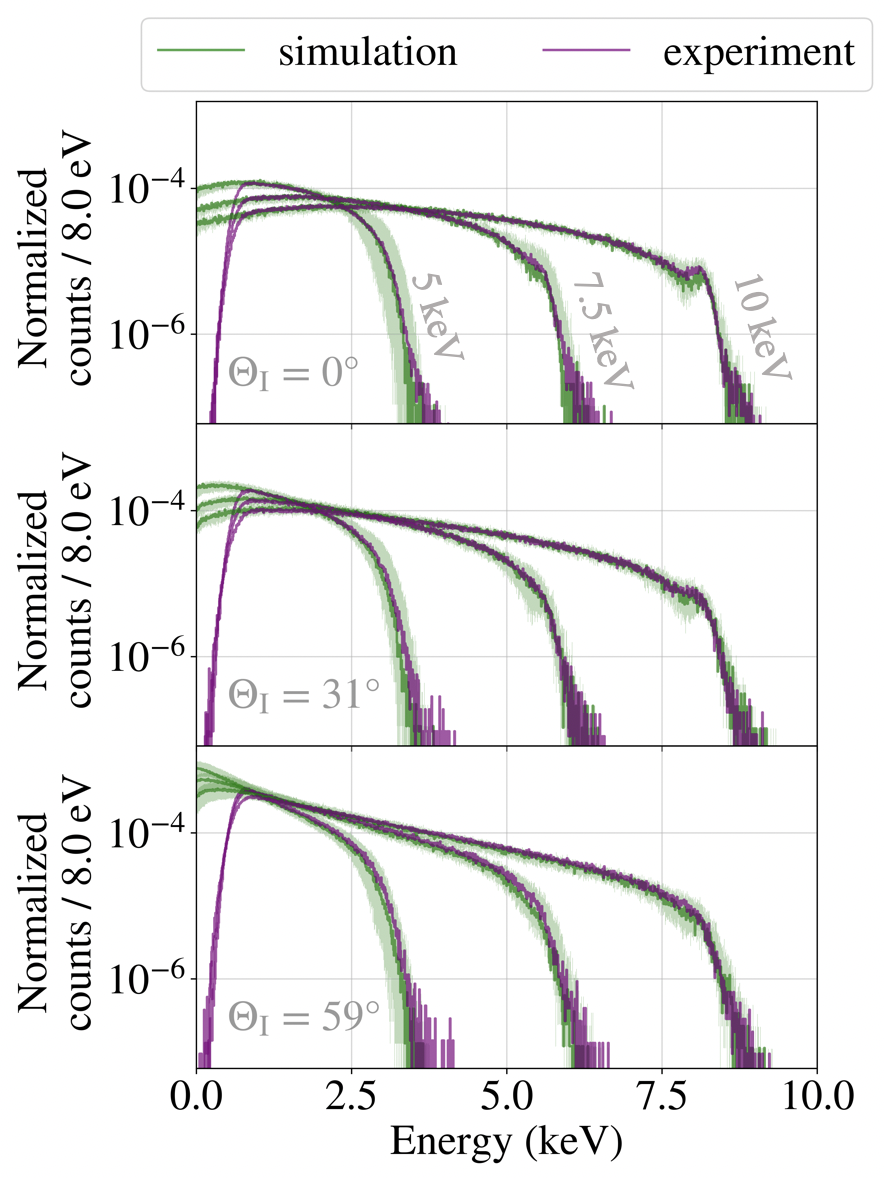}
        \subcaption{Target detector}
        \label{fig:Figure8a}
    \end{subfigure}
    \hfill
    %\qquad
    \begin{subfigure}[b]{0.475\textwidth}
        \centering\includegraphics[width=\textwidth]{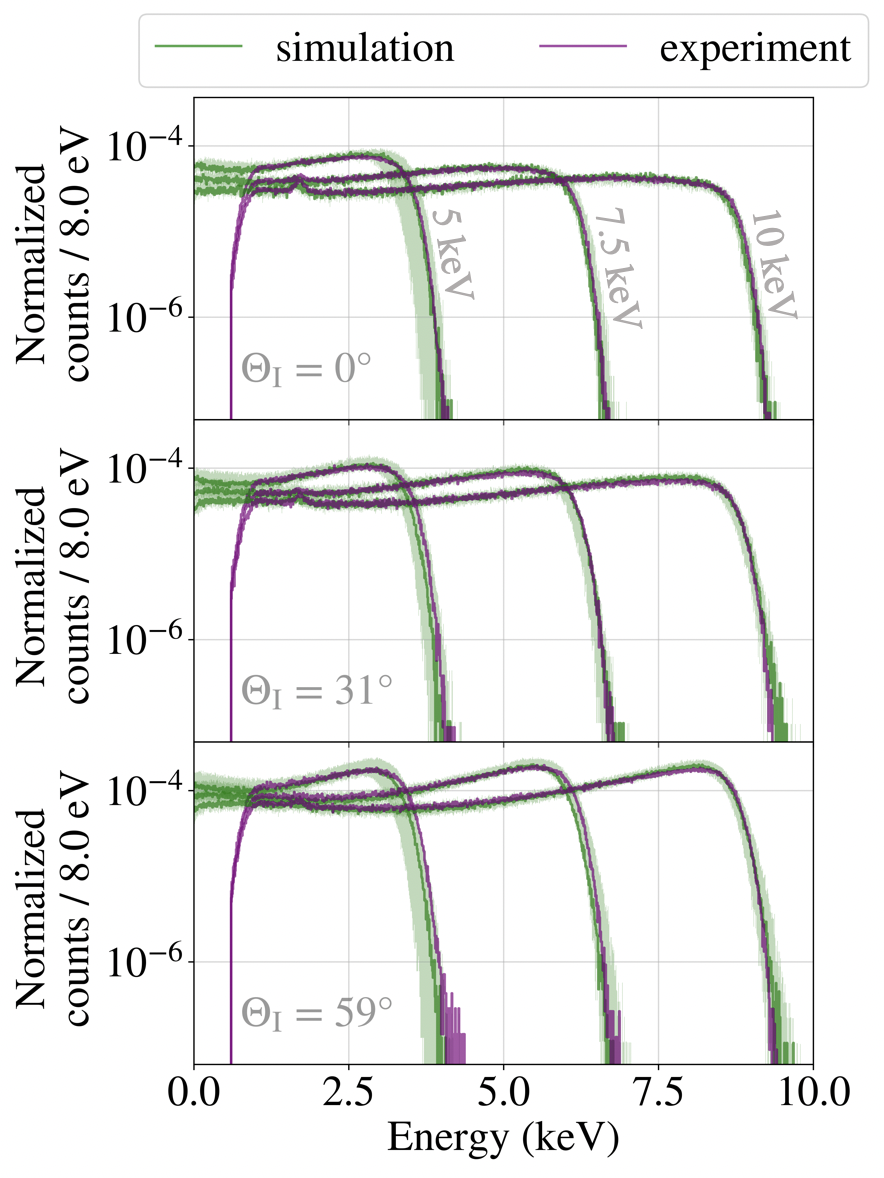}
        \subcaption{Backscattering detector}
        \label{fig:Figure8b}
    \end{subfigure}
\caption{Comparison of coincidence energy spectra. The spectra for coincidence events occurring in the target detector (figure~\ref{fig:Figure8a}) and the backscattering detector (figure~\ref{fig:Figure8b}) are depicted. Each panel shows the spectra for the three different settings of the initial electron energies (\SI{5}{\kilo\electronvolt}, \SI{7.5}{\kilo\electronvolt} and \SI{10}{\kilo\electronvolt}) at a given incident angle. The counts per bin in the spectra are normalized to the total number of recorded electrons in the central pixel of the target detector with energies larger than \SI{0.85}{\kilo\electronvolt} independently for experiment and simulation. For this, the experimentally measured counts are corrected for pileup and charge sharing between pixels. The light green band illustrates the uncertainty introduced in the simulation due to geometrical uncertainties in the experiment and uncertainties of the fitted transition layer parameters.}
\label{fig:Figure8}
\end{figure}

In figure~\ref{fig:Figure8}, the measured energy spectra after the coincident selection and the corresponding simulations are shown for both detectors. The experimental and simulated spectra are in good agreement within the total uncertainty. However, slight discrepancies between the simulation and the experiment can be observed in a closer look at the silicon escape peak. This peak can be observed at \SI{1.74}{\kilo\electronvolt} below the initial electron energy in the spectrum of the target detector. The corresponding photon peak in the backscattering detector is hence positioned at \SI{1.74}{\kilo\electronvolt}. Furthermore, as the effect of the detectors' energy threshold is managed by applying conditions on the minimal energy of coincidence events, as described in section~\ref{sec:Experiment}, incomplete charge collection at the entrance window of the detector majorly defines the rate decrease at $E_\mathrm{I}-E_\mathrm{thres}$. This charge collection efficiency particularly impacts the energy spectra shape for low initial electron energies and high incident angles. In figure~\ref{fig:Figure8}, it can be seen that the discrepancy between the energy spectra of the simulation and the experiment at the truncation at high energies (both in the target and the backscattering detector) is more pronounced, the more influence the charge collection efficiency has. This might hint towards the necessity of a more precise modeling of the charge collection at the entrance window. One other possible explanation for this mismatch could be energy losses due to charge sharing between pixels. Even though the timestamps are utilized to identify charge sharing in the target detector, this identification method is limited due to the energy threshold of the data acquisition system. This problem could be mitigated in future measurements by reducing the electron beam size below the size of a singular pixel such that the charge sharing effects at the pixel border become neglectable.\par 

Multiple simulations are performed using setup two to account for geometric uncertainties in the experiment and inaccuracies in the transition layer parameters used for the simulation. Here, the values for the geometric positioning of the backscattering detector relative to the target detector in three dimensions, the incident and take-off angle, and the transition layer parameters are randomized from a flat distribution of values within the respective uncertainties of each parameter. The uncertainties associated with the geometric parameters are estimated based on the spatial and angular precision achievable within the experimental setup amounting to about \SI{3}{\milli\meter} in position and \SI{3}{\degree} in angle. Uncertainties associated with the transition layer parameters are given in section~\ref{sec:Simulation}. The resulting variation between the simulations is illustrated as an error band in figure~\ref{fig:Figure8}. This method of using multiple simulations with randomized input parameters provides a rough estimation of the total error and the variability of the energy spectra given the uncertainties on various input parameters, which is sufficient for the work presented here.\par 

\FloatBarrier
\subsection{Backscattering coefficients}
\label{sec:ComparisonBScoeff}

The backscattering coefficient is a fundamental parameter quantifying the relation between incoming and backscattered electrons. It is calculated by taking the ratio of electrons detected in the backscattering detector (with a coincidence to an event in the central pixel of the target detector) to those measured in the central pixel of the target detector. To account for the detection efficiencies of both detectors, this ratio is appropriately scaled with the values estimated from simulations. Furthermore, the number of detected incoming electrons in the target detector is corrected by the number of estimated unresolved pileup events and identified charge sharing events for the experimental data.\par

\begin{figure}[tbp]
    \centering
    \includegraphics[width=0.85\textwidth]{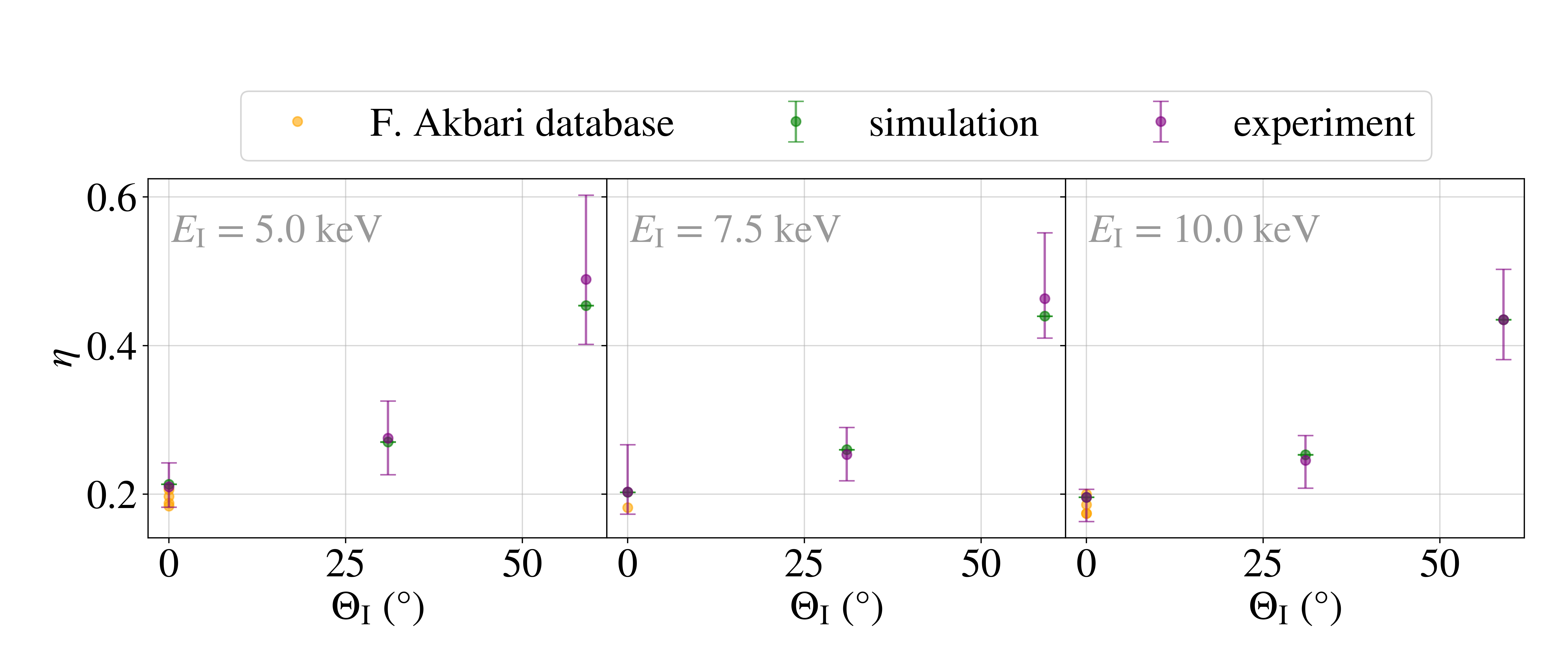}
    \caption{Comparison of backscattering coefficients. In each panel, the ratio of coincidence electrons detected in the backscattering detector to the electrons measured in the central pixel of the target detector is shown as a function of the incident angle at different initial electron energies. The backscattering coefficients of the experiment are corrected for the detection efficiencies of both detectors. For the calculation of the simulated coefficients, simulation setup three is used. Experimentally obtained literature values summarized in the database of F.~Akbari~\cite{Akbari_2023} are provided in orange.}
    \label{fig:Figure9}
\end{figure}

The backscattering coefficient depends on energy and angle, as depicted in figure~\ref{fig:Figure9}. Specifically, there is a significant increase as the incident angle becomes larger. The backscattering coefficient more than doubles, comparing an incident angle of \SI{59}{\degree} with the minimal angle of \SI{0}{\degree}. The backscattering coefficients obtained from the experiment and the simulation are in good agreement within the uncertainties. Additionally, an agreement with literature values was found, which provides confidence in the reliability of the experimental and simulated measurements. For example, a backscattering coefficient of about \SI{19.6}{\percent} for an initial electron energy of \SI{10}{\kilo\electronvolt} and an incident angle of \SI{0}{\degree} is evaluated. This is in good agreement with literature values of around \SIrange{17}{21}{\percent} for silicon materials \cite{Drescher_1970, Joy_1995, Akbari_2023, Renschler_2011}.

The errors on the experimental coefficients, or rather on the detection efficiencies they are scaled with, are roughly estimated by considering the minimum and maximum values from a set of 10 simulations with randomly varied initial geometry and transition layer parameters conducted for each initial energy and incident angle configuration using simulation setup two. For the experiment, statistical uncertainties, as well as uncertainties arising from the estimation of unresolved pileup events and detection of charge sharing events, are negligible. In the simulations, the statistical uncertainty is also negligible, as illustrated in figure~\ref{fig:Figure9}.\par

\FloatBarrier
\section{Conclusions and outlook}

Accurately understanding the effect of electron backscattering on the response of silicon drift detectors (SDDs) is of upmost importance for a keV-scale sterile neutrino search with KATRIN as it highly impacts the sensitivity of the experiment~\cite{Mertens_2021}. For the characterization of this backscattering contribution, an experimental test setup was designed and implemented. It consists of a heated tantalum wire as an electron source and two TRISTAN silicon drift detector devices to detect incoming and backscattered electrons. It was possible to explore the relationships between the energy and angle of incoming and backscattered electrons. Through a coincidence analysis, the contribution of backscattered electrons in the total energy spectrum of the target detector could be extracted.\par 

In a second step, \textsc{Geant4} simulations were used to model the detector response. The experimental and simulation results show a good agreement within the uncertainties considered. Furthermore, the obtained backscattering coefficients, which quantify the ratio of backscattered to incoming electrons, align well with values reported in literature. Despite the large experimental uncertainty, the \textsc{Geant4} simulation has provided a reliable first-order description of electron backscattering. \par

This work will be of major importance for the upcoming keV-sterile neutrino search with the TRISTAN detector. While the results cannot directly be used as an input for the analysis, these measurements and simulations represent an important first step to gauge the possibilities and limitations in quantifying the systematic uncertainties related to backscattering.

The results presented here are among the most precise measurements of the angle and energy distribution of electrons backscattered off silicon. To our knowledge, this effect was investigated for the first time with a tandem system of silicon drift detectors. Hence, other applications of silicon detectors for electron spectroscopy (e.g. in the field of nuclear physics or neutron physics) will also profit from these new results.

Currently, refined \textsc{Geant4} simulations of the full detector response (based on the work of M.~Biassoni~\cite{Biassoni_2021} and A.~Nava~\cite{Nava_2021}), including, for example, charge sharing and data acquisition effects, are being developed for the TRISTAN detector. Furthermore, advanced experimental setups are being commissioned to improve the accuracy of the backscattering measurements. These systems will, among other properties, improve the geometric accuracy and provide more narrow electron beam spots to avoid the effect of charge sharing between pixels.

\acknowledgments
We acknowledge the support of Ministry for Education and Research BMBF (05A20PX3), Max Planck Research Group (MaxPlanck@TUM), Istituto Nazionale di Fisica Nucleare (INFN), Deutsche Forschungsgemeinschaft DFG Graduate School grant no. SFB-1258, and Excellence Cluster ORIGINS in Germany. This project has received funding from the European Research Council (ERC) under the European Union Horizon 2020 research and innovation programme (grant agreement no. 852845).

\bibliographystyle{JHEP}
\bibliography{biblio}

\end{document}